\documentclass[12pt,preprint]{aastex}
\usepackage{times}
\usepackage{natbib}
\usepackage{amsmath}

\newcommand{\nii}{{[N\,{\sc ii}]}}

\newcommand{\hii}{H\,{\sc ii}\rm}

\newcommand{\ha}{{H$\alpha$}}
\newcommand{\oh}{12\,+\,log(O/H)}

\newcommand{\lam}{$\,\lambda$}
\newcommand{\llam}{$\,\lambda\lambda$}
                                                                                
\begin{document}
\title{Metallicity Gradient of a Lensed Face-on Spiral Galaxy at Redshift 1.49}
\author{T.-T. Yuan\altaffilmark{1},  L. J. Kewley\altaffilmark{1}, A. M. Swinbank\altaffilmark{2}, J. Richard\altaffilmark{2,3}, AND R. C. Livermore\altaffilmark{2}}
\altaffiltext{1}{Institute for Astronomy, University of Hawaii, 2680 Woodlawn Drive, Honolulu, HI 96822, USA}
\altaffiltext{2}{Institute for Computational Cosmology, Department of Physics, Durham University, South Road, Durham DH1 3LE, UK}
\altaffiltext{3}{Dark Cosmology Centre, Niels Bohr Institute, University of Copenhagen, Juliane Maries Vej 30, 2100 Copenhagen, Denmark}

\begin{abstract}
We present the first metallicity gradient measurement for a grand-design face-on spiral galaxy at $z\sim1.5$.  
This galaxy has been magnified by a factor of 22$\times$ by a massive, X-ray luminous galaxy cluster MACS\,J1149.5$+$2223 at
$z=0.544$.  Using the Laser Guide Star Adaptive Optics aided integral field spectrograph OSIRIS on KECK II, 
we target the \ha\ emission and achieve a spatial resolution of 0\farcs1, corresponding to a source plane resolution of 170 pc.  The galaxy has well-developed spiral arms and the nebular emission line dynamics clearly indicate a rotationally supported disk with $V_{rot}/ \sigma \sim4$. 
The best-fit disk velocity field model yields a maximum rotation of $V_{rot}{\it \sin{i}}=150\pm15$ km s$^{-1}$,  and a dynamical mass of 
 $M_{dyn}=1.3\pm0.2\times10^{10}{\it csc^2(i)}\,M_{\odot}$ (within 2.5\,kpc), where the inclination angle $i=45\pm10^{\circ}$.  Based on the \nii\ and \ha\ ratios, we measured the radial chemical abundance gradient
 from the inner hundreds of parsecs out to $\sim$5 kpc.  The slope of the gradient is $-0.16\pm0.02$ dex kpc$^{-1}$, significantly steeper than the gradient  of late-type or early-type galaxies in the local universe.   If representative of disk galaxies at $z\sim1.5$, our results support an ``inside-out" disk formation scenario in which early infall/collapse in the galaxy center builds a chemically enriched nucleus, followed by slow enrichment of the disk over the next 9 Gyr. 
  \end{abstract}


\section{Introduction}
The spatial distribution of heavy element abundances is a unique tool to trace the structure formation history of disk galaxies.  Since  the very early work of \citet{Searle71}, \citet{Shields74}, and \citet{Zaritsky94}, it has been well established that there is a negative radial metallicity gradient in spiral galaxies such that the central disk region is more metal-enriched than the outer regions.  The evolution of this gradient over cosmic time provides a powerful constraint on disk formation and evolution models. 
  
Sophisticated chemical evolution models(CEMs) have been successful in reproducing the current gradient of local galaxies, but different models disagree on the time variation of the radial metallicity gradient \citep[e.g.,][]{Molla96,Chiappini97,Magrini07,FuJ09,Marcon10}. Some models predict that the radial gradients steepen with time \citep{Chiappini97,Chiappini01}, while other models predict that the radial gradients flatten with time \citep{Molla97,Hou00,Prantzos00,FuJ09}.  The reason for these opposite predictions is that CEM models are very sensitive to the prescriptions of the detailed physical processes that lead to the chemical enrichment of inner and outer disks.   Observational constraints on these physical
processes are lacking.

Metallicity gradient evolution is unknown due to the small angular sizes and the strong decline of surface brightness of distant galaxies.
Before the next generation of telescopes, the only way to overcome this observational hurdle is to combine the magnification power of gravitational lensing with the Laser Guide Star Adaptive Optics (LGSAO) aided integral field spectrograph units (IFUs) on the world's largest telescopes \citep{Jones10a,Stark08,Swinbank07b}.  
Recently, \citet{Jones10b} reported  the metallicity gradient of a strongly lensed dispersion-dominated galaxy at high-$z$ (``the Clone arc" at $z=2.00$). 
 \citet{Jones10b} measured the slope of the gradient for the inner $\sim$ 1 kpc and found it to be  considerably steeper than local disk galaxies. The steep gradient at high-$z$ 
 is consistent with the ``inside-out" disk formation scenarios. However,  in a low resolution study,  \citet{Cresci10}  reported ``inverted" metallicity gradients for three non-lensed $z\sim3$ Lyman-break galaxies (LBGs), arguing for a ``cold-flow" of primordial gas to the galactic core regions.  Clearly, a large sample of galaxies with chemical abundance gradients across cosmic time are required to place robust constraints on disk formation theories.
 
In this Letter, we report the first metallicity gradient measurement of a lensed {\it spiral} galaxy at $z=1.49$ based on our 
 OH-Suppressing Infra-Red Imaging Spectrograph (OSIRIS) observations at KECK II.  
Throughout this Letter we use a $\Lambda$CDM cosmology with $H_0$= 70 km s$^{-1}$
Mpc$^{-1}$, $\Omega_M$=0.30, and $\Omega_\Lambda$=0.70. At $z=1.5$, 1 arcsec corresponds to 8.5 kpc and a look-back time of 9.3 Gyr. We use solar oxygen abundance 12 + log(O/H)$_{\odot}$=8.66 \citep{Asplund05}.

\section{Observations and Data Reduction}\label{obs}
\begin{figure*}[!ht]
\begin{center}
\vspace{-0.8cm}
\includegraphics[scale=0.56]{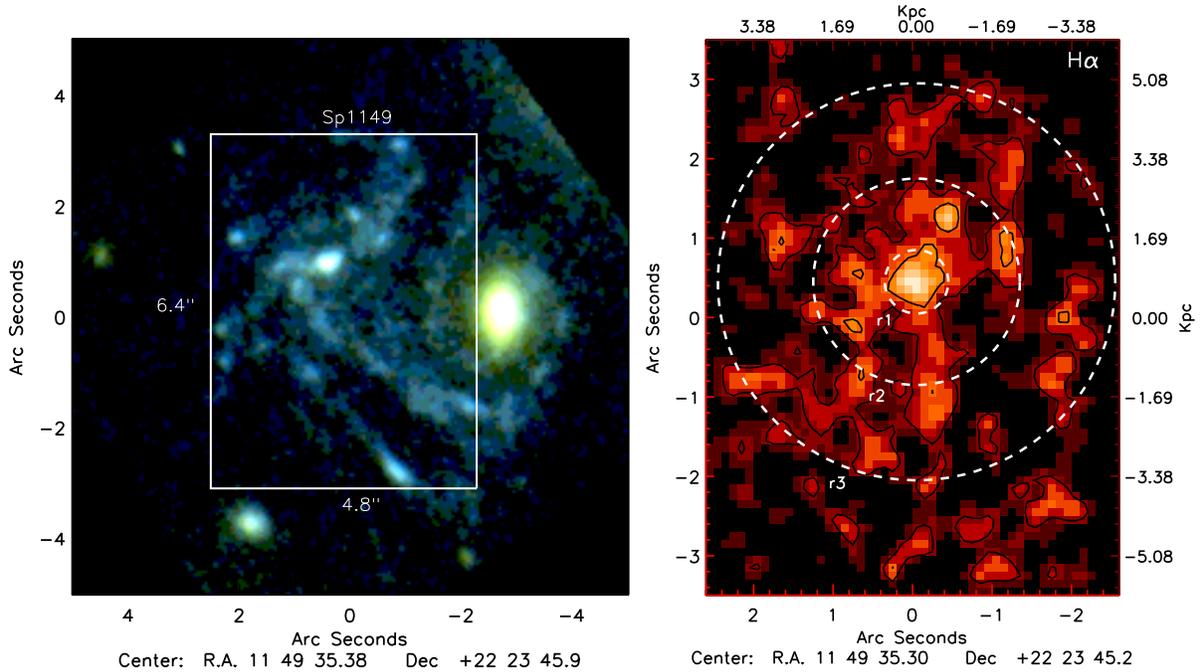}
\caption{Left: HST F814W, F555W-band  color image of Sp1149 behind the lensing cluster MACS\,J1149.5$+$2223.  North is 126$^{\circ}$ counterclockwise to the $y+$ direction. Sp1149 is the largest  of the four multiple images of a spiral galaxy at $z=1.49$.  The white box shows the  
4.$\!\!^{\prime\prime}$8$\times$ 6.$\!\!^{\prime\prime}$4
FOV of OSIRIS Hn3 filter with the 0.$\!\!^{\prime\prime}$1 per pixel plate scale.  Right:  wavelength-collapsed image of the \ha\ emission from our OSIRIS datacube.  Contours in black denote observed \ha\ flux levels of 0.6, 1.3 in units of 10$^{-17}$ ergs~s$^{-1}$~cm$^{-2}$~sec$^{-2}$. Labels on the left and bottom axes indicate the imaging plane coordinates while  the right and top axes show the corresponding physical scale on the source plane.  The white circles mark the three annuli $(r1, r2, r3)$ we use for deriving metallicity gradient (Section~\ref{gradient}).   The \ha\ morphology aligns generally well with the rest-frame UV morphology. 
}
 \label{fig:hst}
\end{center}
\end{figure*}

The spiral galaxy is a four-image lensed system first identified by \citet{Smith09},  located behind the massive, X-ray luminous cluster MACS\,J$1149.5$$+$2223.
We observed the largest  of these four images at $(\alpha_{2000},\delta_{2000})$\,=\,(11$^{\text h}$49$^{\text m}$ 35.$\!\!^{\text s}$284, +22$^{\circ}$23$^{\prime}$45.$\!\!^{\prime\prime}$86) (called Sp1149 hereafter). \citet{Smith09} construct a detailed mass model for the galaxy cluster and show that Sp1149 is 
  one of the largest recorded lensed images of a single galaxy at $z>1$. Owing to the lensing magnification and fortuitous face-on orientation, this galaxy
exhibits more than 10 locally resolved \hii\ regions (Figure~\ref{fig:hst}), ideal for spatial chemical abundance studies. 
The star formation rate from the $V_{555}$ band is  $\sim6~M_{\odot}$ yr$^{-1}$  \citep{Smith09}.

We observed Sp1149 OSIRIS \citep{Larkin06} in conjunction with the LGSAO
system on KECK II telescope on 2010 March 3rd. The weather conditions were moderate with the optical FWHM ranging 
between 0.$\!\!^{\prime\prime}$6 and 1.$\!\!^{\prime\prime}$5.  The average tip/tilt corrected FWHM during the observation was 0.$\!\!^{\prime\prime}$1.   To capture \ha\ and \nii, we used the narrow band Hn3 filter (spectral coverage: 1.594$-$1.676$\micron$; spectral resolution: $\sim$3400), and a plate scale of 0.$\!\!^{\prime\prime}$1 per lenslet. The observations were conducted in  an ABAA dithering sequence with a 10.$\!\!^{\prime\prime}$7 West-North chop onto an object-free sky region.  A total of 19 exposures were obtained for the target,  with 900 s of exposure for each individual frame. The net on-target exposure time is therefore 4.75 hr.  Figure~\ref{fig:hst} shows the OSIRIS field of view (FOV) and locations of Sp1149 on an HST F814W, F555W color image.

Individual exposures were first reduced using the OSIRIS data reduction pipeline \citep{Larkin06}. We used
the sky subtraction IDL code of \citet{Davies07}, which we have modified  to optimize the sky subtraction
specifically around the wavelength region of \ha\ and \nii.  Since we are mainly concerned with the emission lines, 
a first order polynomial function was fit to each spatial sample pixel (spaxel) to improve the removal of lenslet to lenslet variations. The final datacube was constructed by aligning the sub-exposures with the centroid of the \ha\ images and
combing using a 3$\sigma$ mean clip to reject cosmic rays and bad lenslets. Telluric correction and flux calibration was performed by observing a few A0V standard stars immediately after the science exposure. The typical uncertainty in flux calibration is $\sim$10\%.

Gaussian profiles were fitted simultaneously to the three emission lines:  \nii\lam6548, 6583 and \ha. 
The line profile fitting was conducted using a $\chi^2$ minimization procedure which takes into account the greater noise level close to atmospheric OH emissions.
The centroid and velocity width of \nii\llam6548, 6583 lines were constrained by the velocity width of \ha\lam6563, and 
the ratio of \nii\lam6548 and \nii\lam6583 is constrained  to  the theoretical value  \citep{Osterbrock89}. 
An instrumental profile of 5.9\AA\ calculated from sky lines has been subtracted from the line widths.  
We demand  a minimum signal-to-noise (S/N) of 3 to detect each emission line and a minimum S/N of 5 for kinematic and metallicity analysis.

\section{ Kinematics and Lensing Modeling}\label{mor}
To align the astrometry to the IFU datacube, we first assign the brightest pixel of the \ha\ narrow-band
image with the coordinates of the galactic core from  the {\it Hubble Space Telescope (HST)} image, and then rotate the IFU datacube with the observing position angle.  The \ha\ image can 
then be aligned with the {\it HST}  image by matching coordinates. The upper right panel of Figure~\ref{fig:hst} shows our observed \ha\  emission line image, collapsed in wavelength, and smoothed with a Gaussian kernel of FWHM = 2.5 pixels. The \ha\ morphology aligns generally well with the rest-frame NUV/$U$-band image, especially some of the  \hii\ regions in the west.  The SFR and size relation of individual \hii\ regions will be reported in a following paper (R. C. Livermore et al. 2011, in preparation).

We fit the velocity field with rotation disk models as described in \citet{Jones10a}.  
The best fit yields a rotation velocity of   $V_{\text rot} \sin{i}=150\pm15~$km s$^{-1}$, with inclination angle $i=45^{\circ} \pm 10^{\circ}$.
Using $M_{\text dyn}= V_{\text rot}^2\,R / G$, the derived 
dynamical mass is $M_{\text dyn}= 2.6 \pm0.4 \times10^{10} M_{\odot}$ within R=2.5 kpc.
Sp1149 is a rotation dominated disk with $V_{\text rot}$/ $\sigma \sim 4$. 

To reconstruct the source-plane morphology and obtain flux magnification for Sp1149, 
we apply the most recent lens model for cluster MACS\,J1149.5$+$2223 as described in \citet{Smith09}. 
The derived luminosity weighted magnification for Sp1149 is $\mu=22 \pm 2$.  
Accounting for the lensing amplification, the intrinsic magnitude in the  $I-$band is $I\simeq23.4\pm0.3$, corresponding to
$M_B\simeq-20.7$. 

Sp1149 is stretched  $\sim$ 5 on each side $(x, y)$ of the two-dimentional image.  Because of this fortuitous uniform magnification, we can work with the observed datacube and divide a factor of 5 linearly and a factor of 22 in flux to recover the intrinsic linear geometry and luminosity of the galaxy.  Our results are 
therefore unchanged when we use the reconstructed source-plane datacube.

\section{Metallicity Gradient}\label{gradient}
For the metallicity analysis in this work, we use the empirical strong line diagnostic --- the
 N2\,=\,log(\nii\lam6583/\ha) index, as calibrated by \citet[][hereafter PP04]{Pettini04}:
 \begin{equation}
12+{\rm log (O/H)}=8.90+0.57\times {\text N2}
\end{equation}
We  note that the N2 ratio is sensitive to shock excitation, the ionization state of the gas, and 
 the hard ionization radiation field from an active galactic nucleus
 \citep{Kewley02, Kewley06, Rich10}.  Our N2 ratios are consistent with star-formation throughout the disk. The ratio of \nii\,/\ha\ for the whole 
 galaxy is 0.112$\pm$0.050. 
  
\begin{figure}[!ht]
\vspace{-0.8cm}
\begin{center}
\includegraphics[scale=1.]{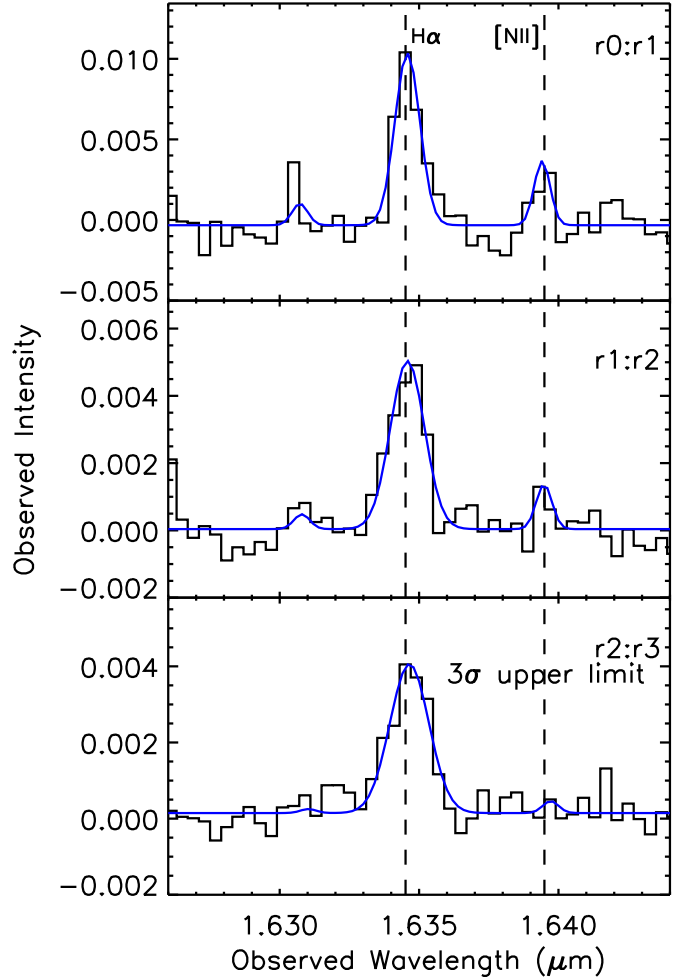}
\caption{Integrated spectra for the three annuli defined in Figure~\ref{fig:hst}.  Vertical dashed lines indicate the   zero-velocity positions  of \nii\ and \ha. The  $y-$axis is the observed intensity in units of  10$^{-17}$ ergs~s$^{-1}$~cm$^{-2}$~\AA$^{-1}$.  The annuli radii $r1, r2, r3$
are chosen to ensure a S/N (\nii)  to be $>$5 for the inner two radii. Blue curves are the Gaussian fits of the spectra. Note that the outer annulus $r2:r3$ is a 3$\sigma$ upper limit for \nii.
The declining ratio of  \nii\,/\ha\  from the central to the outer disk infers a decreasing metallicity with galactocentric distance.
}
\label{fig:spec}
\end{center}
\vspace{-0.4cm}
\end{figure}

Due to its relatively low metallicity (compared with local spirals), the  \nii\ line is too weak to reach a S/N of  5 in individual OSIRIS pixels, but is clearly detected.
We therefore integrate the spectrum within three annuli (see Figure~\ref{fig:hst}, right panel).  The physical lengths of the three outer radii of each annulus in the source plane are  $r1=0.72\pm0.1$ kpc, $r2=2.34\pm0.2$ kpc, $r3=4.5\pm0.4$ kpc.  
The \nii\ line is robustly detected at S/N $>$ 5 for the inner two annuli and is a 3$\sigma$ detection for the outer annulus which we give as an upper limit. The final spectra for the three annuli are presented in Figure~\ref{fig:spec}. 

The metallicities converted from the N2 calibrator for the three annuli are
\begin{align}
 \text{\oh\,}  = \begin{cases}
           8.54\pm0.04            &\text{for} \quad r<r1 \\
        8.38\pm0.05     &\text{for} \quad r1<r<r2 \\
          <8.05                &\text{for} \quad r2<r<r3  \quad (\text{3$\sigma$ upper limit}) \\
         \end{cases} 
\end{align}
 The ``global" metallicity measured from the integrated spectrum of all three regions is 8.36$\pm$0.04. Note that 
the errors of line flux measurements  are smaller than the intrinsic 0.18 dex scatter pertinent to the N2 calibration \citep{Erb06}.
The systematic uncertainty of the N2 calibration with respect to other line diagnostics is about 0.04 dex \citep{Kewley08}.   

A  linear regression with censored data yields a metallicity gradient for the central 4.5 kpc of $\frac{\Delta \log({\rm O/H})}{\Delta {\rm \text R}}= -0.16 \pm 0.02$ dex kpc$^{-1}$, or -0.15 dex kpc$^{-1}$ if we treat the last data point as a measurement rather than an upper limit. The true gradient may be steeper than this value given the 
3$\sigma$ upper limit in the outer annulus. 

\section{Metallicity Gradient Compared with Local and Higher Redshift}
\begin{figure*}[!ht]
\begin{center}
\includegraphics[scale=0.5,angle=0]{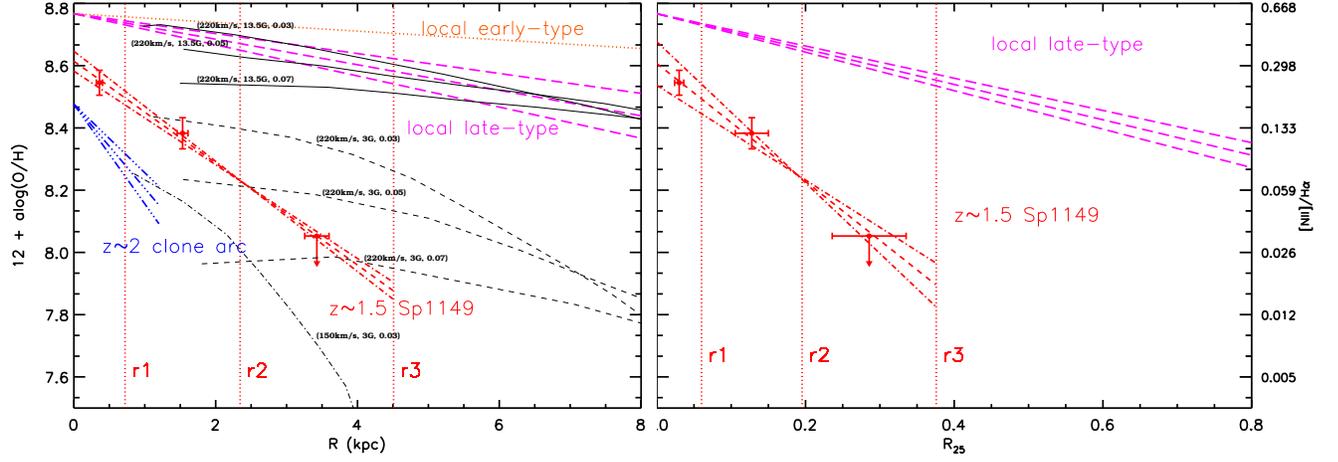}
\caption{
Left: metallicity  vs. galactocentric radius.  
Red lines are the measurements for Sp1149 at z=1.49 from this work. 
The gradient within the central 4.5 kpc is $-0.16\pm0.02$ dex kpc$^{-1}$.  
Vertical red dotted lines show the annulus used to average/sum the spectra.   
Purple dashed lines show the typical gradients of local isolated late-type galaxies, using the control sample of \citet{Rupke10b}. The orange dotted line represents the mean gradient 
of local early-type galaxies, which is typically $\sim$ 3 times shallower than local late-type galaxies \citep{HenryR99}.
Blue lines show the work of \citet{Jones10b} for a dispersion-dominated lensed galaxy  the ``Clone Arc"  at $z=2.0$
who report a even steeper gradient within a size of $\sim$ 1 kpc. 
It is interesting to see that two $z>1$ galaxies have considerably steeper gradients than local galaxies.
Black lines are the model predictions from \citet{Prantzos00} for rotational velocity $V_{c}$=220km~s$^{-1}$ at age t=3Gyr (dashed curves) and $t=13.5$ Gyr (solid curves), with spin parameter $\lambda$=0.03,0.05,0.07.  The black dash-dotted line is the model grid with velocity $V_{c}$=150km~s$^{-1}$ at $t=3$ Gyr and $\lambda$=0.03.  The steep gradients of Sp1149
and ``Clone Arc" are broadly consistent with the shape of the $V_{c}$=150km~s$^{-1}$ model grids.  Right: The same as the left panel, except that the $x-$axis is expressed in  the scaled radius $R_{25}$. 
}
\label{fig:gr}
\end{center}
\vspace{-0.4cm}
\end{figure*}

In Figure~\ref{fig:gr} we compare the metallicity gradient of Sp1149 with galaxies in the local and higher redshift universe, as well as the model predictions from \citet{Prantzos00}.
To avoid systematic effects caused by different calibration indicators, we use the oxygen abundance of \hii\ regions for comparison.    
\citet{Kewley08} showed that metallicities can be compared if the same metallicity calibration is applied consistently across all sample galaxies and \hii\ regions.
We convert all metallicities to the PP04 calibration using the calibrations in \citet{Kewley08}.

\citet{Prantzos00}  modeled metallicity gradients of spiral galaxies at ages $t$\,=\,3, 7.5, and 13.5 Gyr, for different velocities ($V_c$) and spin parameters ($\lambda$).  Since Sp1149 is at an age of $\sim$4.3 Gyr and has $V_c$=212$\pm$50~km~s$^{-1}$, we adopt model grids with $V_c$\,=\,220~km~s$^{-1}$, $t$\,=\,3, 13.5 Gyrs, and $\lambda$\,=\,0.03, 0.05, 0.07 for our comparison.   The model metallicities have been calibrated to the PP04 metallicity diagnostic.  Sp1149 shows a steeper gradient than predicted by \citet{Prantzos00} for $V_c$\,=\,220~km~s$^{-1}$. However, the steep gradient of Sp1149 is broadly consistent with the $V_c$\,=\,150~km~s$^{-1}$ model (the dash-dotted model line in Figure~\ref{fig:gr}).

Metallicity gradients in local galaxies are strongly correlated with Hubble type, bar strength, and merging events; with more flattened gradients for barred galaxies and merging pairs \citep{Kewley10, Rupke10a}.  Therefore it is significant that the metallicity gradient of Sp1149 is substantially steeper than local late-type galaxies  (see Figure~\ref{fig:gr}).
For the local galaxy gradients, we adopt the isolated spiral  control sample of \citet{Rupke10b} comprised of 11 isolated mid-to-late type local galaxies with a mix of bar-strengths.  The median slope for the local control sample is -0.041$\pm$0.009 dex kpc$^{-1}$, with a standard deviation of 0.03 dex kpc$^{-1}$.  
The median slope of gradients from the Rupke et al. (2010) sample is consistent with the values of \citet{Zaritsky94} and \citet{vanZee98} samples. 
The \citet{Zaritsky94}  gradient sample composed of 39 local disk galaxies, and the \citet{vanZee98} sample added another 11 local spirals.
Local early-type galaxies have even shallower gradients \citep{HenryR99}. 

A similar (and even steeper) gradient has been reported in a recent work of \citet{Jones10b} for  a 
 dispersion-dominated lensed system within only $\sim$1 kpc at $z$=2.
  \citet{Cresci10} recently reported reversed metallicity gradients (i.e., the core is less enriched than the outer disk) for three un-lensed LBGs at $z$$\sim$ 3. They explained the inverted gradients by inflow of pristine cold gas to the galactic center.   Due to their low resolution ($\sim$ 4-6 kpc),  the LBG selection, and potential presence of  mergers and galactic winds which may contaminate metallicity gradient,  it is not clear how the \citet{Cresci10} results relate to the work presented here.

In the right panel of Figure~\ref{fig:gr} we show that the metallicity gradient of Sp1149 is still considerably steeper than local galaxies when expressed in units of dex/($R$/$R$$_{25}$), 
 where $R_{25}$ is the $B-$band isophotal radius at a surface brightness of 25 mag arcsec$^{-2}$.  
 We estimate the  $R_{25}$ for Sp1149 by fitting the isophotal profile of the source-plane F814W-band image assuming 
 a color difference of $U-B= -0.1$ mag  for Sc type galaxies  \citep{Fukugita95}.  The resulting $R_{25}$ for Sp1149 is 12$\pm$2kpc, comparable to the range of $R_{25}$ 
 in our local control sample \citep{Kewley10,Rupke10b}.

\begin{figure}[!ht]
\begin{center}
\includegraphics[scale=0.6,angle=90]{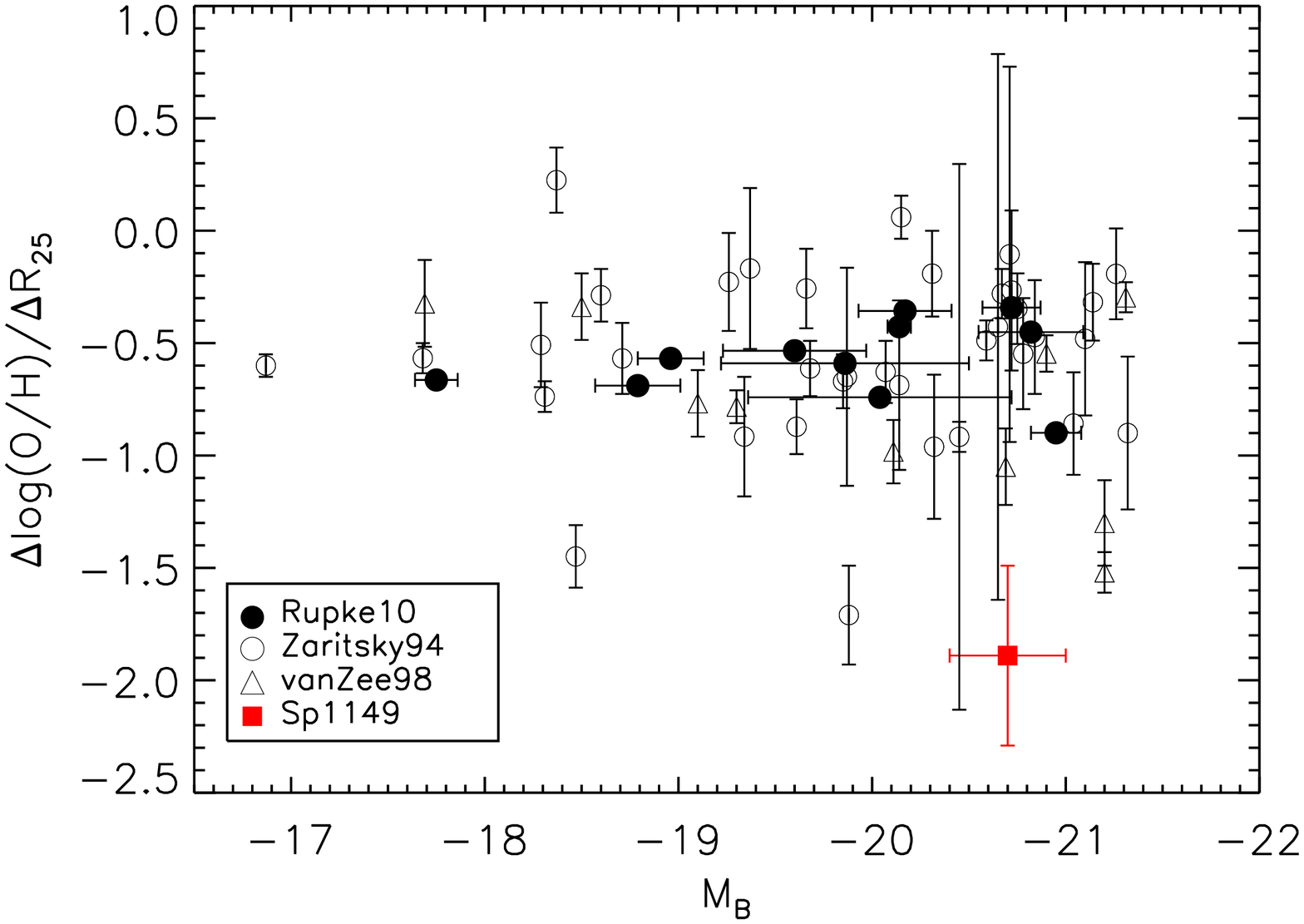}
\caption{Metallicity  gradient (dex/$R_{25}$) vs absolute B band magnitude $M_{B}$.  
The red square represents the measurement  for Sp1149 at $z$=1.49.
Black filled circles indicate the locations of the local late-type galaxies of \citet{Rupke10b}.
Black empty circles are from the sample of \citet{Zaritsky94}, where the metallicity gradients are measured for 39 local disk galaxies.
Black empty triangles come from the sample of \citet{vanZee98}, where the metallicity gradients are provided for another 11 local spiral galaxies.
Sp1149 has significantly steeper gradient than local galaxies of comparable luminosity, even when expressed in units of scaled radius.
}
\label{fig:grdiag}
\end{center}
\vspace{-0.4cm}

\end{figure}

To analyze the effect of size and luminosity on the metallicity gradients, 
in Figure~\ref{fig:grdiag}, we plot the metallicity gradients (in units of $R_{25}$) v.s. the absolute $B-$band magnitude $M_B$ for local late-type galaxies and Sp1149.
Sp1149 is comparable to the brightest local late-type disks, while ``the Clone Arc" is much more luminous ($M_B=-22.12$).  We see from Figure~\ref{fig:grdiag} that 
the scaled gradient of Sp1149 is steeper than in typical local galaxies. 

\section{Discussion}\label{discuss}
The redshift range of $z$=1---3 is the period when star-formation, mass assembly and chemical enrichment activity peaks \citep[e.g.,][]{Fan01,Dickinson03, Chapman05,HopkinsAM06,Conselice07}. Metallicity gradient studies of galaxies in this redshift range therefore give important insights into disk galaxy formation.  The picture that the metallicity gradient flattens with cosmic time is qualitatively consistent with 
the ``inside-out" disk formation scenario \citep[e.g.,][]{Prantzos00, Bouwens97,Benson10}.  
This scenario predicts that in the early stages of 
galaxy evolution,  the inner galactic bulge undergoes a rapid collapse with vigorous star formation at the center,
building a steep radial metallicity gradient from the core to the outer disk.   Violent gas dynamics and accretion events marked these early stages of disk formation. Subsequent radial mixing  and infall cause the gradient to flatten over the following Gyrs, resulting in a weaker 
 gradient on average in late-type galaxies today.   

In this work, we have presented an OSIRIS IFU study of a grand-design lensed spiral galaxy at $z=1.49$. The 22$\times$ magnification allows us to resolve individual \hii\ regions  at an intrinsic resolution of 170 parsec. We measure the radial distribution of chemical abundances in three annuli and find that the metallicity decreases from 
a central value of 12+{\rm log (O/H)}=8.54 to less than 8.05 in the outer disk. The metallicity slope is $-0.16\pm0.02$ dex kpc$^{-1}$ for the inner 4.5 kpc, much steeper than that of late-type galaxies of comparable luminosities in the local universe, even when expressed in $R_{25}$ units.  Our results are broadly consistent with the ``inside-out" model of disk formation.

Metallicity gradient observations for a larger number of star-forming galaxies at different redshifts are required  to build a well-defined sample to form a solid observational picture of  the metallicity gradient evolution as a function of cosmic time.  Before the next generation of telescopes, we are aiming to secure a few tens of gravitationally lensed systems 
as a pioneer sample to study the gradient evolution of galaxies at high-z.   Future instruments such as the {\it JWST}, TMT and GMT will improve the current resolution limits by $\sim$
1 order of magnitude, revolutionizing chemical gradient evolution studies at high-$z$.

\acknowledgments 
We thank the referee for his/her valuable comments which have helped a lot to improve this Letter.
This work is based on data obtained at the W. M. Keck
 Observatory. We are grateful to the Keck Observatory staff for assistance with our observations, especially Jim Lyke, Hien Tran, and 
Randy Campbell.   T.-T.-Y. thanks the hospitality of ICC, 
Durham University, where a large portion of the work was done. 
A.-M.-S. acknowledges a STFC Advanced Fellowship. R.-C.-L. acknowledges a studentship from STFC.
The authors recognize the very significant cultural role that the summit of Mauna Kea has  within the indigenous Hawaiian community.  


\end{document}